\documentclass{article}
\usepackage{amssymb}
\usepackage{bm}
\usepackage[dvips]{graphicx}
\begin{document}

\title{Current-induced resonance\\ in a ferromagnet--antiferromagnet junction}

\author{S. G. Chigarev$^1$, E. M. Epshtein$^1$\thanks{E-mail: epshtein36@mail.ru}, Yu. V.
Gulyaev$^1$,\\
V. D. Kotov$^1$, G. M. Mikhailov$^2$, P. E. Zilberman$^1$
\\ \\
$^1$V. A. Kotelnikov Institute of Radio Engineering and Electronics\\
of the Russian Academy of Sciences, Fryazino, 141190, Russia\\ \\
$^2$Institute of Microelectronics Technology and High Purity Materials\\
of the Russian Academy of Sciences, Chernogolovka, 142432, Russia}

\date{}

\maketitle

\abstract{We calculate the response of a ferromagnet--antiferromagnet
junction to a high-frequency magnetic field as a function of the
spin-polarized current through the junction. Conditions are choused under
which the response is zero in absence of such a current. It is shown that
increasing in the current density leads to proportional increase in the
resonance frequency and resonant absorption. A principal possibility is
indicated of using ferromagnet--antiferromagnet junction as a terahertz
radiation detector.}

\section{Introduction}\label{section1}
On a level with ``conventional'' spintronics studying effects in
ferromagnet/ferromagnet junctions, a new direction has emerged which is related
with spin-polarized current effect on the antiferromagnetic layer in
ferromagnet/antiferromagnet junction~\cite{Nunez}--\cite{Gulyaev3}, so
that a term ``antiferromagnetic spintronics'' has
appeared~\cite{Nunez,Shick}.

\begin{figure}
\includegraphics{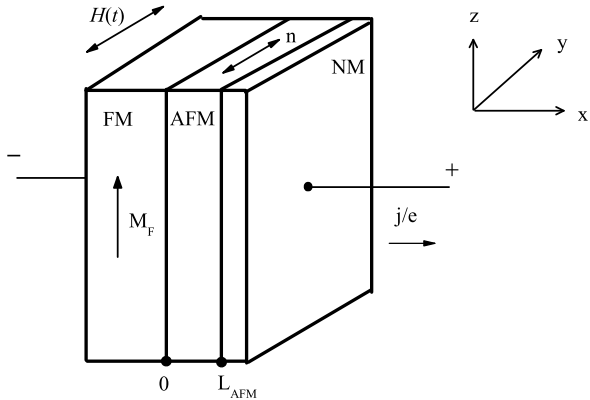}
  \caption{Scheme of the ferromagnet (FM)--antiferromagnet (AFM)
  junction; NM being a nonmagnetic layer. The main vector directions are
  shown.}\label{fig1}
\end{figure}

The interest in studying magnetic junctions with antiferromagnetic layers
is related with the following features of antiferromagnets. First, this is
low, compared to ferromagnets, magnetization in magnetic fields much lower
than the exchange field. This allows to neglect demagnetization effect
and, that is more significant, leads to substantially lower values of
magnetic fields and currents at which switching effects occur. In contrast
with ferromagnets, where spontaneous magnetization exists even in absence
of magnetic fields and currents, so that the role of the latter consists
in changing the magnetization direction, in the antiferromagnets with
mutually compensated magnetic sublattices existence of the resulting
magnetic moment is due to magnetic fields and/or currents. Second, the
eigenfrequency of the magnetic oscillation in antiferromagnets exceeds
the similar frequency in ferromagnets by several orders of magnitude, so
that the range of possible using of antiferromagnetic structures extends
up to terahertz (THz) frequencies.

One of the directions in spintronics is studying the spin-polarized current effect
on ferromagnetic resonance in magnetic
junctions~\cite{Fuchs}--\cite{Okutomi}. The current effect on the spectrum
and damping of the magnetization oscillation in antiferromagnets was
studied in Refs.~\cite{Gulyaev2,Gulyaev3}. It was shown that the
spin-polarized current effect leads to decrease in damping and to instability
of the antiparallel configuration with switching to parallel one. A
possibility was noted of creating canted  antiferromagnet configuration
with appearance of net magnetization under spin-polarized electron
injection without external magnetic field.

In present article, we consider forced oscillation of the antiferromagnet
magnetization under high-frequency magnetic field in presence of
spin-polarized current. The conditions are choused under which the response
to the high-frequency is zero when such a current is absent. Under such
conditions, the current exerts a strong effect on the antiferromagnet
resonant characteristics. Such a current-driven resonator may be used, in
principle, as a THz detector.

\section{The model and basic equations}\label{section2}
Let us consider a magnetic junction consisting of a pinned ferromagnetic
(FM) layer, a free antiferromagnetic (AFM) layer, and nonmagnetic (NM)
layer closing the electric circuit (Fig.~\ref{fig1}). A thin spacer is
supposed between FM and AFM layers to prevent exchange interaction through
the interface. The current flows perpendicular to the layers (along $x$
axis) in the direction corresponding to electron flux from ferromagnet to
antiferromagnet. The easy axis of the antiferromagnet lies in the layer
plane (along $y$ axis), the ferromagnet magnetization vector $\mathbf M_F$
is parallel to $z$ axis.

The AFM layer thickness $L_{AFM}$ is assumed to be small compared
to the spin diffusion length, so that the macrospin approximation is
valid. In this approximation, the layer magnetization is supposed to be
uniform in thickness, while the spin current through the interface is
taking into account by means of additional terms in the equations (see
Refs.~\cite{Gulyaev4,Gulyaev2} for details). A simplest AFM model is
considered with two collinear equivalent sublattices with equal
magnetizations, $|\mathbf M_1|=|\mathbf M_2|=M_0$.

The Landau--Lifshitz--Gilbert equations for the AFM layer in the presence
of spin-polarized current and high frequency magnetic field take the following
form (see detailed derivation in Ref.~\cite{Gulyaev2}):
\begin{eqnarray}\label{1}
  &&\frac{d\mathbf M}{dt}-\frac{1}{2}\frac{\kappa}{M_0}
    \left\{\left[\mathbf M\times\frac{d\mathbf M}{dt}\right]
    +\left[\mathbf L\times\frac{d\mathbf L}{dt}\right]\right\}
    +\gamma\left[\mathbf M\times\mathbf H(t)\right]\nonumber \\
    &&+\frac{1}{2}\gamma(\beta+\beta')(\mathbf M\cdot\mathbf n)[\mathbf M\times\mathbf
    n]+\frac{1}{2}\gamma(\beta-\beta')(\mathbf L\cdot\mathbf n)[\mathbf L\times\mathbf
    n]\nonumber \\
    &&+K\left[\mathbf M\times\left[\mathbf M\times\hat{\mathbf
    M}_F\right]\right]+P\left[\mathbf M\times\hat{\mathbf
    M}_F\right]=0,
\end{eqnarray}
\begin{eqnarray}\label{2}
  &&\frac{d\mathbf L}{dt}-\frac{1}{2}\frac{\kappa}{M_0}
    \left\{\left[\mathbf L\times\frac{d\mathbf M}{dt}\right]
    +\left[\mathbf M\times\frac{d\mathbf L}{dt}\right]\right\} \nonumber \\
    &&+\gamma\left[\mathbf L\times\mathbf H(t)\right]
    -\gamma\Lambda\left[\mathbf L\times\mathbf M\right] \nonumber \\
    &&+\frac{1}{2}\gamma(\beta+\beta')(\mathbf M\cdot\mathbf n)[\mathbf L\times\mathbf
    n]+\frac{1}{2}\gamma(\beta-\beta')(\mathbf L\cdot\mathbf n)[\mathbf M\times\mathbf
    n] \nonumber \\
    &&+K\left[\mathbf L\times\left[\mathbf M\times\hat{\mathbf
    M}_F\right]\right]+P\left[\mathbf L\times\hat{\mathbf M}_F\right]=0.
\end{eqnarray}
Here, $\mathbf M=\mathbf M_1+\mathbf M_2$ is the AFM magnetization vector,
$\mathbf L=\mathbf M_1-\mathbf M_2$ is the antiferromagnetism vector,
$\hat{\mathbf M}_F$ is the unit vector along the FM layer magnetization,
$\mathbf n$ is the unit vector along the anisotropy axis, $\mathbf{H}(t)$
is the external magnetic field, $\kappa$ is the damping coefficient,
$\Lambda$ is the uniform exchange constant, $\beta,\,\beta'$ are the
intrasublattice and intersublattice anisotropy constants, respectively ($\beta>\beta'$ is
assumed), $\gamma$ is the gyromagnetic ratio,
\begin{equation}\label{3}
  K=\frac{\mu_BQ}{eL_{AFM}M^2}j,
\end{equation}
\begin{equation}\label{4}
  P=\frac{\gamma\alpha_{sd}\mu_B\tau Q}{eL_{AFM}}j,
\end{equation}
$j$ is the current density, $\mu_B$ is the Bohr magneton, $\alpha_{sd}$ is
the (dimensionless) \emph{sd} exchange interaction constant, $\tau$ is the
spin relaxation time in the AFM layer, $Q$ is the FM conductivity spin
polarization, $e$ is the electron charge.

The last two terms in the right-hand sides of Eqs.~(\ref{1}) and~(\ref{2})
describe (in the macrospin approximation) the spin-polarized current
effect on the antiferromagnet magnetic configuration. There are two
mechanisms of this effect. One of them~\cite{Slonczewski,Berger} is due to
relaxation of the noncollinear (with respect to the AFM magnetization)
component of the electron spins with transfer corresponding torque to the
lattice. This occurs within a distance comparable with the Fermi
wavelength from the FM--AFM interface. The injected spins collinear to AFM
magnetization with lost transverse component remain in nonequilibrium
state within much longer distance of the order of the spin diffusion length.
Such a state is energetically unfavorable. This can lead to change of the
lattice magnetic configuration with transition to more favorable state.
This is the second mechanism of the spin-polarized electron interaction
with magnetic lattice~\cite{Heide,Gulyaev5}. These mechanisms are
described by the terms with $K$ and $P$ coefficients, respectively. It is
seen from Eqs.~(\ref{1}) and~(\ref{2}) that the second mechanism is
equivalent to the presence of an additional magnetic field $P\hat{\mathbf
M}_F/\gamma$ parallel to the FM magnetization vector. The latter
circumstance leads to appearance of a current-induced AFM canted state
in absence of external magnetic field.

In the configuration described, $\hat{\mathbf M}_F=\{0,\,0,\,1\},\;\mathbf
n=\{0,\,1,\,0\}$. It is suggested that external dc magnetic field is
absent, while a high-frequency magnetic field is parallel to the
anisotropy axis, $\mathbf H(t)=\{0,\,H_0\cos\omega t,\,0\}$ with $H_0\ll
H_E$, where $H_E\equiv\Lambda M_0$ is the exchange field. Under such
conditions, the AFM magnetization is zero in absence of the current
($j=0$). Correspondingly, the magnetic susceptibility component
$\chi_{yy}$ responsible for absorption of the high-frequency field with
the polarization indicated~\cite{Akhiezer} is zero, too, so that AFM
resonance does not occur.

\section{Spin-polarized current-driven resonance in antiferromagnet}\label{section2}
As it was mentioned, antiferromagnet magnetization appears along the
ferromagnet magnetization vector $\hat{\mathbf M}_F$ under spin-polarized
current. Precession of the antiferromagnet magnetization vector makes
possible the resonance absorption.

Let us calculate the antiferromagnet magnetization with using
Eqs.~(\ref{1}) and~(\ref{2}). The high-frequency field is assumed to be
low, is taken into account in scope of the linear approximation, and,
hence, does not influence the static magnetization, which is~\cite{Gulyaev2}
\begin{equation}\label{5}
  \overline{\mathbf
  M}=\{0,\,0,\,\overline{M}_z\},\quad\overline{M}_z=\frac{P}{\gamma\left(\Lambda+
  \frac{1}{2}(\beta-\beta')\right)}\approx\frac{P}{\gamma\Lambda}.
\end{equation}
The corresponding antiferromagnetism vector is
\begin{equation}\label{6}
  \overline{\mathbf
  L}=\{0,\,\overline{L}_y,\,0\},\quad\overline{L}_y=\sqrt{4M_0^2-\overline{M}_z^2}.
\end{equation}
To calculate the response to the (low) high-frequency field, we linearize Eqs.~(\ref{1}),
(\ref{2}) in small deviations from $\overline{\mathbf M},\,\overline{\mathbf
L}$. The following set of equations is obtained:
\begin{eqnarray}\label{7}
  &&\frac{\partial M_x}{\partial t}-\frac{1}{2}\frac{\kappa}{M_0}
  \left\{-\overline M_z\frac{\partial M_y}{\partial t}+
  \overline L_y\frac{\partial L_z}{\partial t}\right\}+PM_y
  \nonumber \\
  &&-\frac{1}{2}\gamma(\beta+\beta')\overline M_zM_y
  -\frac{1}{2}\gamma(\beta-\beta')\overline L_yL_z+K\overline M_zM_x\nonumber\\
  &&=\gamma\overline M_zH(t),
\end{eqnarray}

\begin{equation}\label{8}
  \frac{\partial M_y}{\partial t}-\frac{1}{2}\frac{\kappa}{M_0}
  \overline M_z\frac{\partial M_x}{\partial t}-
  PM_x+K\overline M_zM_y=0,
\end{equation}

\begin{equation}\label{9}
  \frac{\partial\widetilde M_z}{\partial t}+\frac{1}{2}\frac{\kappa}{M_0}
  \overline L_y\frac{\partial L_x}{\partial
  t}+\frac{1}{2}\gamma(\beta-\beta')\overline L_yL_x=0,
\end{equation}

\begin{eqnarray}\label{10}
  &&\frac{\partial L_x}{\partial t}-\frac{1}{2}\frac{\kappa}{M_0}
  \left\{\overline L_y\frac{\partial\widetilde M_z}{\partial t}-
  \overline M_z\frac{\partial\widetilde L_y}{\partial t}\right\}\nonumber\\
  &&-\gamma\left\{\Lambda+\frac{1}{2}(\beta-\beta')\right\}\overline L_y
  \widetilde M_z=0,
\end{eqnarray}

\begin{equation}\label{11}
  \frac{\partial\widetilde L_y}{\partial t}-\frac{1}{2}\frac{\kappa}{M_0}
  \overline M_z\frac{\partial L_x}{\partial
  t}-\frac{1}{2}\gamma(\beta-\beta')\overline M_zL_x=0,
\end{equation}

\begin{equation}\label{12}
  \frac{\partial L_z}{\partial t}+\frac{1}{2}\frac{\kappa}{M_0}
  \overline L_y\frac{\partial M_x}{\partial t}+
  \gamma\left\{\Lambda+\frac{1}{2}(\beta-\beta')\right\}\overline L_yM_x
  +K\overline M_zL_z=0,
\end{equation}
where $\widetilde M_z=M_z-\overline M_z,\;\widetilde L_y=L_y-\overline L_y
$.

We seek a solution in the form of forced oscillation with frequency $\omega$
of the external magnetic field. We find
\begin{equation}\label{13}
  M_x(\omega)=-\frac{\left(-i\omega+P/\eta\right)PH_0}{\Lambda\left(\omega^2-
  \Omega^2+2i\nu\omega\right)}\equiv\chi_{xy}(\omega)H_0,
\end{equation}
\begin{equation}\label{14}
  M_y(\omega)=-\frac{P^2H_0}{\Lambda\left(\omega^2-
  \Omega^2+2i\nu\omega\right)}\equiv\chi_{yy}(\omega)H_0,
\end{equation}
\begin{equation}\label{15}
  \Omega^2=2\gamma^2H_AH_E+P^2+\left(\frac{\gamma H_E}{\eta}\right)^2,
\end{equation}
\begin{equation}\label{16}
  \nu=\gamma H_E\left(\kappa+\frac{1}{\eta}\right),
\end{equation}
where $H_A=(\beta-\beta')M_0$ is the anisotropy field, $\eta=\alpha\gamma
M_0\tau$.

Absorption of the high-frequency field is determined by the imaginary part
of the diagonal susceptibility
\begin{equation}\label{17}
  \chi''_\|\equiv\mathrm{Im}\chi_{yy}(\omega)
  =\frac{2\nu\omega P^2}{\Lambda\left[\left(\omega^2-\Omega^2\right)^2+4\nu^2\omega^2\right]}.
\end{equation}
The maximal absorption corresponds to the resonance frequency
\begin{equation}\label{18}
  \omega_\mathrm{res}=\sqrt{\Omega^2-2\nu^2}.
\end{equation}
The Q-factor of the system is
\begin{equation}\label{19}
  Q=\frac{\Omega}{2\nu}.
\end{equation}
It follows from Eqs.~(\ref{15}) and~(\ref{16}) that the resonance
frequency and Q-factor rise under increase in current.

The power absorbed in a unit volume is~\cite{Akhiezer}
\begin{equation}\label{20}
  W=\frac{1}{2}\omega\chi''_\|H_0^2,
\end{equation}
while the linear absorption coefficient for an electromagnetic wave
incident on the layer is
\begin{equation}\label{21}
  \Gamma=\frac{8\pi W}{cH_0^2}=4\pi q\chi''_\|,
\end{equation}
where $c$ is the light velocity, $q=\omega/c$ is the wavenumber of the
incident electromagnetic wave.

\begin{figure}
\includegraphics{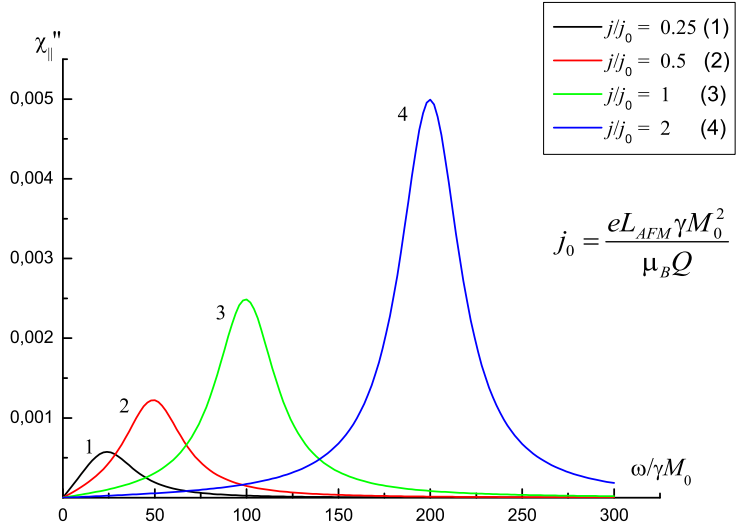}
  \caption{Imaginary part of the diagonal susceptibility $\chi''_\|$ as a
  function of the dimensionless frequency $\omega/\gamma M_0$ under
  various (dimensionless) current densities $j/j_0$.}\label{fig2}
\end{figure}

\section{Discussion}\label{section4}
Let us make numerical estimates using the following parameter values:
$M_0\sim10^3$ G, $\Lambda\sim10^3$, $\alpha\sim 10^4$,
$\beta\sim\beta'\sim10^{-1}$, $\kappa\sim10^{-2}$, $\tau\sim10^{-12}$ s,
$L_{AFM}\sim10^{-6}$ cm. We find $H_E\sim10^6$ G, $H_A\sim10^2$ G,
$\nu\sim10^{11}$ s$^{-1}$, $\Omega\sim10^{11}$ s$^{-1}$, $\eta\sim10^2$.
As a scale of the current density, we choose the quantity
\begin{equation}\label{22}
  j_0=\frac{eL_{AFM}\gamma M_0^2}{\mu_BQ_F},
\end{equation}
so that
\begin{equation}\label{23}
  P=\eta\gamma M_0\frac{j}{j_0}.
\end{equation}
With indicated parameter values, $j_0\sim10^7$ A/cm$^2$. At $j\sim j_0$ we
have $\Omega\approx P$, i.e., the eigenfrequency is proportional to the
current density. The same applies to the resonant absorption.

The absorption spectrum ($\chi''_\|$ as a function of the dimensionless
frequency $\omega/\gamma M_0$) with various current densities is shown in
Fig.~\ref{fig2}. It is seen, that the resonance frequency and resonant
absorption rise proportionally to the current density. At $j\sim j_0$, we
have the resonance frequency about $10^{12}$ c$^{-1}$, that corresponds to
THz range.

\begin{figure}
\includegraphics{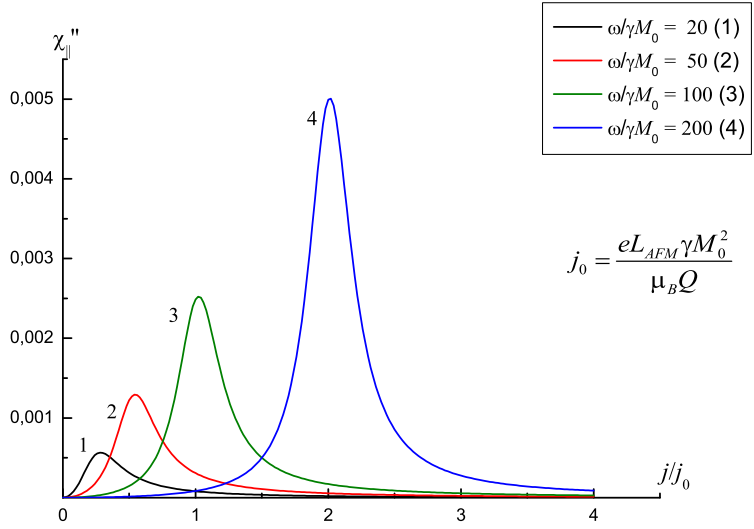}
  \caption{Imaginary part of the diagonal susceptibility $\chi''_\|$ as a
  function of the dimensionless current density $j/j_0$ at various
  (dimensionless) frequencies $\omega/\gamma M_0$.}\label{fig3}
\end{figure}

The absorption as a function of the current density at various frequencies
has the similar form (see Fig.~\ref{fig3} where the same dimensionless
variables are used).

At $\Omega=10^{12}$ s$^{-1},\,\nu=10^{11}$ s$^{-1}$ we have $Q=5$. (For
comparison: the Q-factor of free oscillation without current
\begin{equation}\label{24}
  Q_0=\frac{1}{\kappa}\sqrt{\frac{H_A}{2H_E}}
\end{equation}
is less than 1.) The Q-factor rises under increase in the frequency and/or
current density.

For THz radiation ($q\sim10^2$ cm$^{-1}$) at $j\sim j_0$ the absorption
coefficient is $\Gamma\sim10$ cm$^{-1}$, so that the absorption within
the thickness of the AFM layer is quite small, $\sim10^{-5}$--$10^{-4}$. To overcome
this difficulty, a multilayer structure of alternating
ferromagnet--antiferromagnet layers may be used with electromagnetic wave
incident from the butt side (along $z$ axis). The reflection coefficient
of the normally incident wave is~\cite{Sokolov}
\begin{equation}\label{25}
  R=\left|\frac{\widetilde n-\widetilde\mu}{\widetilde n+\widetilde\mu}\right|^2,
\end{equation}
where $\widetilde n=n+ik$ is the complex index of
refraction, $\widetilde\mu=\mu'+i\mu''=1+4\pi(\chi'+i\chi'')$ is the
complex magnetic permeability.

\begin{figure}
\includegraphics{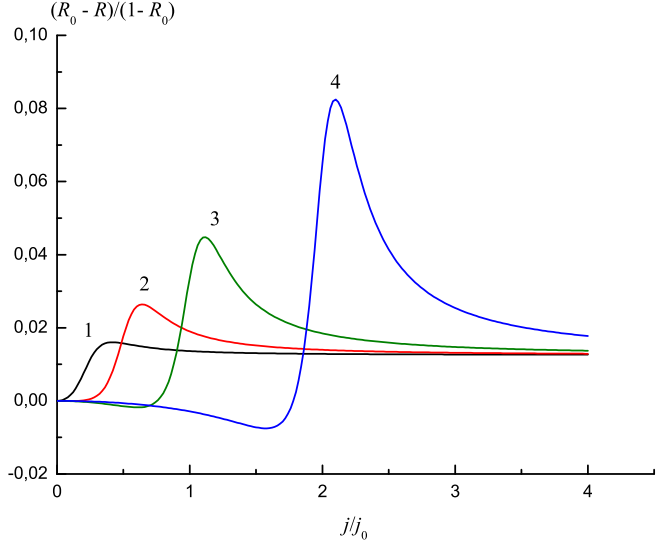}
  \caption{The current-induced relative change of the reflection
coefficient as a function of the (dimensionless) current density. The
notations are the same as in Fig.~\ref{fig3}.}\label{fig4}
\end{figure}

In long wavelength range~\cite{Sokolov}
\begin{equation}\label{26}
  n\approx k\approx\sqrt{\frac{2\pi\sigma_0}{\omega}}\gg1,\,|\mu|,
\end{equation}
where $\sigma_0$ is the static conductivity. Therefore, Eq.~(\ref{25}) can
written as
\begin{equation}\label{27}
  R=R_0-4\pi(1-R_0)(\chi'+\chi''),
\end{equation}
where $R_0$ is the reflection coefficient in absence of the spin-polarized
current when $\chi'=\chi''=0$ under geometry in consideration (see
Eq.~(\ref{14})).
The current-induced relative change of the reflection
coefficient $(R_0-R)/(1-R_0)=4\pi(\chi'+\chi'')$ as a function of the (dimensionless)
current density is shown in Fig.~\ref{fig4}. It is seen that $\Delta
R=R-R_0$ is of the order of $10^{-3}$--$10^{-2}$ at $j\sim j_0$ (i.\,e.
$\sim10^7$ A/cm$^2$ at chosen parameters). The desired resonance signal
can be extracted by the current modulation.

\section{Conclusion}\label{section5}
The results indicate a possibility of a new effect, namely,
current-induced resonance in ferromagnet--antiferromagnet
junctions. The resonance frequency and
resonant absorption are proportional to the current density through the
junction. This opens a principal possibility of using such junctions as
current-controlled resonant detectors for THz radiation. Making of
corresponding experiments seems to be interesting.\\

The work was supported by the Russian Foundation for Basic Research, Grant No.
10-02-00030-a.


\begin{thebibliography}{19}

\bibitem{Nunez}
A.S. N\'u\~nez, R.A. Duine, P. Haney, A.H. MacDonald, Phys. Rev. B \textbf{73}, 214426 (2006).
\bibitem{Wei1}
Z. Wei, A. Sharma, A.S. Nunez, P.M. Haney, R.A. Duine, J.Bass, A.H. MacDonald,
M. Tsoi, Phys. Rev. Lett. \textbf{98}, 116603 (2007).
\bibitem{Urazhdin}
S. Urazhdin, N. Anthony, Phys. Rev. Lett. \textbf{99}, 046602 (2007).
\bibitem{Haney}
P.M. Haney, R.A. Duine, A.S. N\'u\~nez, A.H. MacDonald, J. Magn. Magn. Mater. \textbf{320}, 1300 (2008).
\bibitem{Gomonay1}
H. Gomonay, V. Loktev, J. Magn. Soc. Jpn. \textbf{32}, 535 (2008).
\bibitem{Wei2}
Z. Wei, A. Sharma, J. Bass, M. Tsoi, J. Appl. Phys. \textbf{105}, 07D113 (2009).
\bibitem{Gomonay2}
H.V. Gomonay, V.M. Loktev, Phys. Rev. B \textbf{81}, 144127 (2010).
\bibitem{Shick}
A.B. Shick, B. Khmelevskyi, O.N. Mryasov, J. Wunderlich, T. Jungwirth, Phys. Rev. B \textbf{81}, 212409 (2010).
\bibitem{Hals}
K.M.D. Hals, Y. Tserkovnyak, A. Brataas, Phys. Rev. Lett. \textbf{106}, 107206 (2011).
\bibitem{Gomonay3}
H.V. Gomonay, R.V. Kunitsyn, V.M. Loktev, arXiv:1106.4231.v2 [cond-mat.mtrl-sci].
\bibitem{Gulyaev1}
Yu.V. Gulyaev, P.E. Zilberman, E.M. Epshtein, J. Commun. Technol. Electron. \textbf{56}, 863
(2011).
\bibitem{Linder}
J. Linder, Phys. Rev. B \textbf{84}, 094404 (2011).
\bibitem{Gulyaev2}
Yu.V. Gulyaev, P.E. Zilberman, E.M. Epshtein, J. Exp. Theor. Phys. \textbf{114}, 296 (2012).
\bibitem{Gulyaev3}
Yu.V. Gulyaev, P.E. Zilberman, E.M. Epshtein, J. Commun. Technol. Electron. \textbf{57},
(2012) (to be published).
\bibitem{Fuchs}
G.D. Fuchs, J.C. Sankey, V.S. Pribiag, L. Qian, P.M. Braganca, A.G.F. Garcia,
E.M. Ryan, Zhi-Pan Li, O. Ozatay, D.C. Ralph, R.A. Buhrman, Appl. Phys. Lett. \textbf{91}, 062507 (2007).
\bibitem{Petit}
S. Petit, N. de Mestier, C. Baraduc, C. Thirion, Y. Liu, M. Li, P. Wang, B.
Dieny, Phys. Rev. B \textbf{78}, 184420 (2008).
\bibitem{Posth}
O. Posth, N. Reckers, R. Meckenstock, G. Dumpich, J. Lindner, J. Phys. D: Appl. Phys.
\textbf{42}, 035003 (2009).
\bibitem{Boone}
C.T. Boone, J.A. Katine, J.R. Childress, V. Tiberkevich, A. Slavin, J. Zhu, X. Cheng, I. N.
Krivorotov, Phys. Rev. Lett. \textbf{103}, 167601 (2009).
\bibitem{Seki}
T. Seki, H. Tomita, A.A. Tulapurkar, M. Shiraishi, T. Shinjo, Y. Suzuki, Appl. Phys. Lett. \textbf{94}, 212505
(2009).
\bibitem{Guan}
Y. Guan, J.Z. Sun, X. Jiang, R. Moriya, L. Gao, S.S.P. Parkin, Appl. Phys. Lett. \textbf{95}, 082506
(2009).
\bibitem{Chen}
W. Chen, G. de Loubens, J.-M.L. Beaujour, J.Z. Sun, A.D. Kent, Appl. Phys. Lett. \textbf{95}, 172513
(2009).
\bibitem{Wang}
R.-X. Wang, P.-B. He, Q.-H. Liu, Z.-D. Li, A.-L. Pan, B.-S. Zou, Y.-G. Wang,
J. Magn. Magn. Mater. \textbf{322}, 2264 (2010).
\bibitem{Staudacher}
T. Staudacher, M. Tsoi, Thin Solid Films \textbf{519}, 8260 (2011).
\bibitem{Okutomi}
Y. Okutomi, K. Miyake, M. Doi, H.N. Fuke, H. Iwasaki, M. Sahashi, J. Appl. Phys. \textbf{109}, 07C727
(2011).
\bibitem{Gulyaev4}
Yu.V. Gulyaev, P.E. Zilberman, A.I. Panas, E.M. Epshtein, J. Exp. Theor.
Phys. \textbf{107}, 1027 (2008).
\bibitem{Slonczewski}
J.C. Slonczewski, J. Magn. Magn. Mater. \textbf{159}, L1 (1996).
\bibitem{Berger}
L. Berger, Phys. Rev. B \textbf{54}, 9353 (1996).
\bibitem{Heide}
C. Heide, P.E. Zilberman, R.J. Elliott, Phys. Rev. B \textbf{63}, 064424 (2001).
\bibitem{Gulyaev5}
Yu.V. Gulyaev, P.E. Zilberman, E.M. Epshtein, R.J. Elliott, JETP Lett. \textbf{76}, 155 (2002).
\bibitem{Akhiezer}
A.I. Akhiezer, V.G. Baryakhtar, S.V. Peletminskii, Spin Waves, North-Holland Publ. Co., Amsterdam,
1968.
\bibitem{Sokolov}
A.V. Sokolov, Optical Properties of Metals, American Elsevier Publ. Co.,
1967.

\end{thebibliography}
\end{document}